\documentclass{IEEEtran}
\usepackage{cite}
\usepackage{amsmath,amssymb,amsfonts}
\usepackage{algorithmic}
\usepackage{graphicx}
\usepackage{textcomp}
\def\BibTeX{{\rm B\kern-.05em{\sc i\kern-.025em b}\kern-.08em
    T\kern-.1667em\lower.7ex\hbox{E}\kern-.125emX}}
\begin{document}

\title{Unified Pandemic Tracking System Based on Open Geospatial Consortium SensorThings API}

\author{
	\IEEEauthorblockN{Robinson~Paniagua \IEEEauthorrefmark{1}, Radwa~Sultan \IEEEauthorrefmark{1}, and
	Ahmed Refaey\IEEEauthorrefmark{1}}

				\IEEEauthorblockA{\IEEEauthorrefmark{1}Electrical and Computer Engineering Department, Manhattan College, New York, USA}}




\maketitle
\begin{abstract}
With the current nations struggling to track the pandemic's trajectories. There has been a lack of transparency or real-live data streaming for pandemic cases and symptoms. This phenomenon has led to a rapid and uncontrolled spread of these deadly pandemics. One of the main issues in creating a global pandemic tracking system is the lack of standardization of communications protocols and the deployment of Internet-of-Things (IoT) device sensors. The Open Geospatial Consortium (OGC) has developed several sensor web Enablement standards that allow the expeditious deployment of communications protocols within IoT devices and other sensor devices like the  OGC  SensorThings application programming interface (API). In this paper, to address this issue,  we outline the interoperability challenge and provide a qualitative and quantitative study of the OGC SensorThings API's deployment and its respective server. The OGC SensorThings API is developed to provide data exchange services between sensors and their observations. The OGC SensorThings API would play a primary and essential role in creating an automated pandemic tracking system. This API would reduce the deployment of any set of sensors and provide real-time data tracking. Accordingly, global health organizations would react expeditiously and concentrate their efforts on high infection rates. 
\end{abstract}

\maketitle

\section{Introduction}
After the recent spread of COVID-19 (coronavirus) that has emerged as the Genesis of the deadliest viruses\cite{1}, Governments have been taking reactionary measures to curb the spread of this infectious illness\cite{2}. Global pandemics have been the Genesis of catastrophic effects and immeasurable stressors within global healthcare systems and global economics. Many factors can accelerate the pandemic spread and lead to an exponential rise in the number of cases \cite{3}:

\begin{itemize}
  \item [--] Lack of Transparency - Delayed dissemination of knowledge
  \item [--] Delayed Travel Restrictions - Aviation services lacked insufficient screening measures
  \item [--] Public Misinformation - Dissemination of false or inaccurate information, leading to confusion and conspiracy theories.
  \item [--] Announcement Delays - Delays of public health organizations declaring a state of emergencies due to lack of information
  \item [--] Global leaders not adhering to transparent reporting will under-report COVID-19 cases. This will further propagate the spread of the virus because appropriate responses are directly dependent on the accuracy of reporting.
\end{itemize}

These factors all have the same resonating tone or issue. This issue is due to some communication breakdown, delays in the reporting mechanism, or lack of transparency. The reporting mechanism for such outbreaks cannot be relied upon through verbal or mechanical means. Accordingly, an effective automated virus tracking system is crucially needed. Proper tracking, or perhaps automatic tracking and reporting, is essential in deploying timely and effective countermeasures such as geographical containment. Currently, Pandemic tracking has been dependent on inefficient reporting strategies.  

To mitigate any future pandemics, the reporting mechanisms must migrate to an automatic solution using Internet of Things (IoT) sensor devices and cloud technologies. The global arena must implement and obtain live data streaming and on-demand data collection through a tracking network that contains proactive detection and monitoring sensors that automatically send the readings to a centralized database. A simple proactive detection system could include infrared temperature sensors. These infrared sensors are already being used to provide temperature readings and could be a good indicator that a person might be infected and be a carrier of the virus. These sensors would be situated strategically along with street corners and checkpoints. These checkpoints can include but are not limited to places of mass gatherings, train station terminals, buses, and airports. Once the sensor detects the abnormal body temperature, the sensor can obtain these metrics and simultaneously send the readings to the centralized server or database. Afterward, these alerts could be forwarded to the proper medical authorities for further action as tracking the propagation of the virus or perhaps preventing that person from further spreading the virus. These live data collections will enable expeditious medical response and containment of any geographically infected areas indicated by the sensors \cite{x1,x2,x3,x4}.

However, designing the network architecture for implementing live data collection and streaming creates inherent challenges. These challenges are intrinsic to every aspect of the sensor network architecture. These challenges are as follows
 \begin{itemize}
 \item [--] Applications / Presentation Protocol  - the architecture will need an application that receives these metrics and applies Data Science to these metrics. The application will need to categorize between abnormal and normal readings, insert these readings into a graphical user interface, and send a message to the proper personnel. Within this application, Machine Learning algorithms are better suited to handle this workload.
 \item [--] Transport Protocol - to implement a highly scalable network (perhaps thousands of sensors within a geographical area), protocols such as TCP will cause latency and congestion issues within the network. The sensors' communication system must implement protocols, such as UDP, which are specifically designed for speed and congestion control.
 \item [--]  Networking - the architecture must lend itself to a highly scalable and efficient transport solution. In high-density population cities, such as New York City, if the volume of data transmission and reception is not implemented efficiently, it will overwhelm the network. Proper segmentation and IP scheme formatting within the network must be of utmost importance.
 \item [--] Data Transmission  - the architecture must transmit the data in a lightweight format to prevent congestion and minimize latency.
 \item [--] Network Management - due to the complex nature of this network architecture, networking devices such as switches, routers, firewalls, and web authentication portals, the network managers must implement device-level management tools for configuration changes and implementation of QoS and network parameters.
 \item [--] Vendor Dependency - to fulfill rapid deployment and global utilization of these sensor networks, the network devices must not be limited to one specific vendor. This in itself also creates other communications and logistical issues. Vendors will use different frequencies, communications protocols, security protocols, QoS settings, and configuration requirements.
 \item [--] Data Storage and processing - the amount of data collected will be astronomical, perhaps within the Tetra-byte range per geographic region. The network must categorize and store this data for analytical purposes and graphical representation.
 \item [--] Security - the network must address security protocols such as intrusion detection, intrusion prevention, authentication, prevention of viruses/malware, and periodic security scans and updates.
\end{itemize}

The current healthcare system has effectively deployed IoT devices to track and monitor certain illnesses and provide immediate assistance according to the signal provided by these devices. In the research conducted in \cite{5}, authors propose the use of a Multi-access edge computing IoT system and hosting the multi-access edge computing (MEC)-enabled IoT system using Open Geospatial Consortium standards with the application of an ad hoc single-purpose application program interface (API). When deployed on a global scale, these devices must conform to the current 802.11ah standards. The standards indicate that every country has its own proprietary frequency in which these sensors would operate. That would elevate the cost of mass-producing any IoT sensor with the goal of worldwide deployment \cite{x5,x6,x7,x8}.

\begin{figure*}
    \centering
    \includegraphics[height = 13cm, width=15cm]{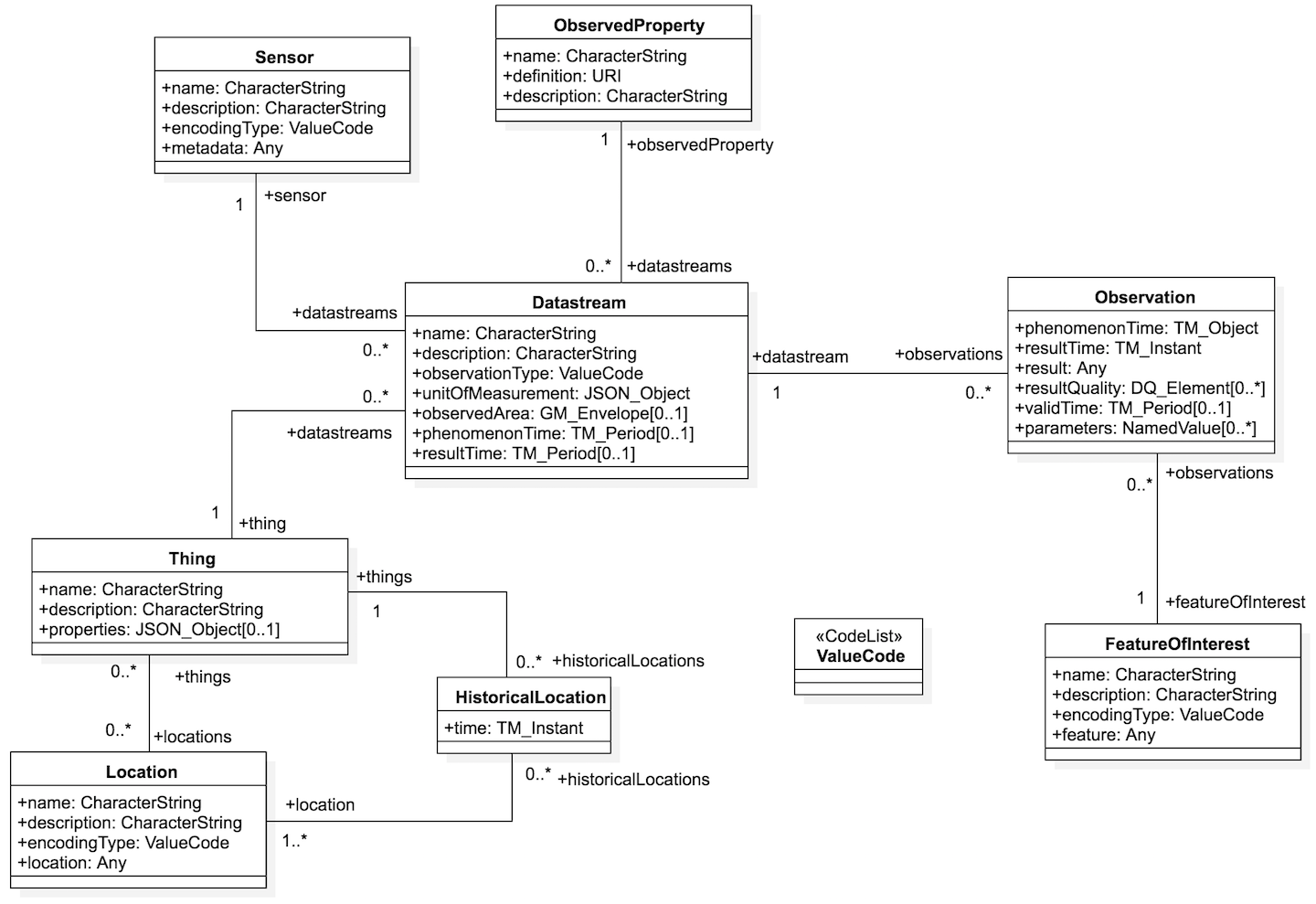}
    \caption{OGC Sensor API Model \cite{5}}
    \label{fig:UML}
\end{figure*}

However, Analysis of IoT data is critical but without the expense of resource depletion. As shown in Fig. \ref{fig:UML} which illustrates the proposed IoT Sensor Eco-System Architecture in \cite{5}, the authors propose using an alignment engine technique in which the relationship of instances within a collected dataset mapping the data to a joint latent space reducing bandwidth of transmission. The system proposed in \cite{5} receives data at the edge level and samples the sensor compilations on an edge server. The edge server will apply the appropriate filtering using the proposed hyper-alignment engine and forward it to the Cloud. The hyper-alignment engine amalgamates sensor data from users in varying geographic locations, regions, and time zones. These compilations of data can ultimately consist of body temperature readings and any other symptoms experienced. 

While the proposed Eco-system provides a good basis for collecting data, there are a few inherent challenges with this eco-system. 
 \begin{itemize}
 \item [--] The end device communicates through WiFi /Bluetooth transceivers. Some cities are equipped with Metropolitan Area Networks (MAN), providing WiFi capabilities across the city to virtually every device within its range. While this is evident in cities like New York City, many other global regions or even cities within the US do not provide such services. 
 \item [--]  The WiFi/Bluetooth communication protocol lends itself to lost data streams and poor connection. In a densely populated environment or city, such as New York City or in any given instance, there are millions of users operating on the same ISM band, on which WiFi/Bluetooth protocol operates. Accordingly, the reliability (via RF interference), throughput, and speed at which the edge devices transmit their data will be substantially reduced. 
 \item [--]  The WiFi/Bluetooth architecture insufficiently allows for scaling of the network due to range. WiFi/Bluetooth are homogeneous networks and typically employ a point-to-multi-point topology. A MESH topology would better serve the network in terms of reliability, efficiency, and throughput.
 \item [--]  Finally, the system considers transmitting OGC SensorThings API at the edge level. The performance of this type of architecture can be potentially hindered when communicating with the main server. The edge-level device is dependent on  HTTP traffic to the server from each device deployed. Considering a global arena, the number of sensors needed to materialize the tracking system would be substantial. Therefore, individual HTTP connections with all these devices can potentially lead to congestion and possible data loss \cite{11}\cite{x9,x10,x11,x12}. In addition, it is assumed that the geographical placement of the sensor will have access to the Internet. In regions where the internet is insufficient, UDP / multicast can be used via MESH network radios. These radios can extend coverage from one access point within the MANET. This idea will be explored in upcoming subsections of this paper.
\end{itemize}

With the advent of MIMO technology and the advancement of MESH ad hoc networking radios, these radios have the potential to solve range, reliability, scalability, edge computing, and interference issues created by the ISM band. This technology uses a multitude of transport systems to its intended endpoint. One point-to-point device can exceed communication ranges by more than 10 - 30 miles depending on terrain. When these devices are placed into a MESH network, the intended endpoint can no longer have a line of sight from the originating radio or sensor. This architecture is essential in geographical regions that lack or are depleted of  LTE / Cellular and WiFi infrastructure. This paper further discusses the implications of using this type of network in later sections.

Furthermore, the authors in \cite{5} propose utilizing cloud services that implement real-time monitoring and observation. The cloud service will aggregate all the data and create a graphical interface for the consumer. These cloud services are indeed an essential and critical aspect of translating the data into a graphically consumable product. There are cloud services currently in existence that perform such operations. The Geospatial Open Consortium Sensorofthings API can conjugate all this data and produce a graphical representation.

\indent On the other hand, mobile applications are another unprecedented and viable method that can be developed in an attempt to track the symptoms and trajectory of pandemics. Throughout recent years, there have been many mobile applications and sensors dedicated to monitoring certain diseases. Among these diseases are cardiovascular disease, cancer, HIV, diabetes, bipolar disorder, and other critical illnesses \cite{4}. Although these applications can successfully track certain diseases, using a mobile application to track a pandemic contains several inherent deficiencies. Additionally, due to the rapid and exponential spread of pandemics, there has not been a globally accepted monitoring standard. In theory, mobile application development for self-reporting symptoms provides a comprehensive method for tracking the pandemic and providing guidance or assistance to the patient. This application can also collect metrics and aid in the strategic collaboration for curbing the pandemic. Many of these applications are not centralized, have their reporting standards, and only report to their subscribers or user base.

In this paper, we provide an analysis of combining all these sensor technologies and communications protocols to establish pandemic monitoring and tracking architecture. The proposed architecture provides a standard in which a universal communications protocol is implemented as an API. This API will serve to send patient data to a unified library. This virtual repository will serve as a global focal point for data exchange \cite{x13,x14,x15}. This architecture would allow for any participant irrespective of geolocation to automatically report any symptoms or cases. 

The remainder of this paper is organized as follows. In Section \ref{tab:sec_open} the standardization of the communications protocols and the component breakdown of the OGC API are discussed. In Section \ref{tab:sec_ogc} a comparative analysis of the potential use of the OGC API as communications protocols and standards for a global IoT network is presented. The paper culminates with conclusions and recommendations based on the comparative analysis.

\section{Open Geospatial Consortium SensorThings API}\label{tab:sec_open}
At the apex of the COVID-19 network tracking system resides the virtual unified data repository. The sole function of the virtual unified repository consolidate all sensor data irrespective of the originating device, communications protocols, vendor, or geolocation. The virtual unified data repository is an essential and critical element for establishing seamless communications and graphical interpretation of global sensor data. Interpreting and utilizing global sensor data becomes cumbersome and inefficient due to the varying specificity of each vendor's sensor construction and communications platform. Rather than "reinvent the wheel" and invest in a costly framework, the Open Geospatial Consortium has created a virtual platform for the sole purpose of seamlessly interconnecting IoT devices across the web.

The OGC SensorThings API is primarily derived and fundamentally built from the  OGC Sensor Web Enablement (SWE) standards. These standards are aimed at interconnecting sensor devices and providing the ISO/OGC Observation and measurement data model [OGC 10-004r3 and ISO 19156:2011]. The key intent of the OGC SensorThings API is to provide the community with an open and geo-spatially enabled unified standard to interconnect IoT mechanisms. These mechanisms include data exchange between non-proprietary sensors, web applications, and devices. These standards had one common goal establishing a structure for exchanging sensor data and metadata. The OGC SWE suite, including its derivative and SensorThings API, utilizes machine-readable language such as Sensor Model Language (SensorML), Observation and Measurement (OM), and Geography Markup Language (GML). These machine-readable language encodings are efficiently used to produce and describe observations, from their respective sensors, which are ultimately loaded for sharing \cite{x16,x17}, analysis, and exploration on the World Wide Web.

The fundamental function of the OGC SensorofThings API is to establish "relational connections between entities" of varying systems to create models of the real world. The SensorThings API allows IoT sensing devices from all aspects of the industry to collaborate and communicate under one common hierarchy. This API was specifically designed to prevail in resource-constrained devices, hardware, or sensors. 

Additionally, the OGC SensorofThings API should provide the essential and critical virtual framework that provides a junction that acts as a unifying data exchange mechanism for heterogeneous devices with any geographic location or region. This virtual framework is also needed for the fusion of the network data and the graphical interpretation of the results as it will not require standardizing the edge or core devices to any specific hardware or protocol. 

\begin{figure*}
    \centering
    \includegraphics[height = 10cm, width=10cm]{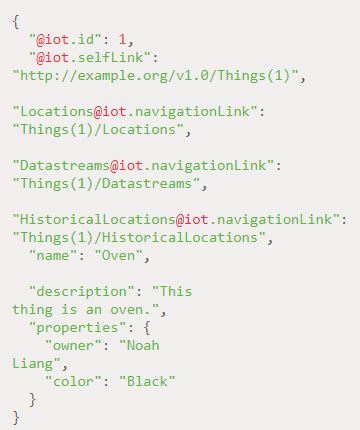}
    \caption{OGC Sensor API Thing Entity}
    \label{fig:Entity}
\end{figure*}

Accordingly, that virtual platform has solved many complexities that involve standardizing and globalization of an interwoven pandemic tracking network. The issues residing at various networking and communication levels, as mentioned above, can be solved by implementing the OGC SensorofThings (SAT) API. The OGC SAT API addresses critical intercommunication requirements such as:

\begin{itemize}
    \item [--] Standardizing of communications protocols across all platforms.
    \item [--] Providing an application layer utility that can be represented through the OGC's graphical interface or a graphical interface specific to the consumer's preference.
    \item [--] Setting forth a standard for posting, retrieving, deleting, and editing data.
    \item [--] Establishing a lightweight transmission protocol for transferring data from an edge device.
    \item [--] Creating a centralized, real-time, transparent virtual repository of ongoing tracking events.
    \item [--] Eliminating any one vendor monopolizing the network with vendor-specific equipment.
    \item [--] Creating a transparent environment in which any consumer can download the data and analyze source information.
    \item [--] Diversifying the edge devices to be a multitude of varying devices, such as sensors, and mobile and desktop applications for self-reporting.
\end{itemize}

The OGC SensorThings API is primarily categorized under two cohesive entities. These entities are called Sensing and Tasking. In the proposed work, our main scope is the Sensing entity of the OGC SensorThings API, which enables a standardized method of retrieving observations and metadata collected from the sensor. It also provides efficient data management and transfer protocol between heterogeneous IoT sensor systems \cite{21}. The Tasking entity is a work in progress with the intent of tasking sensors or actuators (those that are able) with parameters or tasks dependant on observed data working in conjunction with the Sensing entity.

The OGC SensorThings API uses the most commonly known and widely used REST (Representational State Transfer). The OGC API is based on HTTP communications protocols to interconnect IoT devices and services via the WEB using the JSON (JavaScript Object Notation) architectural standards. The RESTful  API architecture enables a systematic standard that is widely used, known, and implemented throughout the web. It is the most commonly known website inter-connectivity architecture known to web developers and enables the ease of altering, navigating, filtering, and data customization.

The OGC data model is directly derived from the ISO observation and Measurement conceptual model. The model consists of ten entity types and their relationships.

\begin{itemize}
    \item \textbf{Thing}: The thing can be the COVID monitoring station or the people that are being monitored.
    \item \textbf{Location}: The location defines the geographical area or region where the things are situated.
    \item \textbf{Historical Location}: Things move. If the location of the thing changes, it will store past and present locations.
    \item \textbf{Sensor}: The sensor is the electronic device that creates the observation.
    \item \textbf{Observed Property}: The result of the observation is actuated by the electronic device or devices.
    \item \textbf{Data Stream}: The data stream interconnects the set of observations of the Observed Property and stores the unit of measurement.
    \item \textbf{Feature of Interest}: Gives the Observation a set location.
    \item \textbf{Observation/phenomenon Time}: The exact time of the occurrence or sensor observation. The time can be an instant, time interval or an average of times in which the observations took place.
    \item \textbf{Observation/result}: The result is the numeric or values of the observation taken by the sensor.
    \item \textbf{Thing/properties}: These are JSON object parameters in which the user can store pertinent data within these fields. These fields are also searchable.
\end{itemize}

The OGC SensorThings API is formatted using the Javascript Open Notation (JSON) language. This language uses human-readable text for exchanging data. By using JSON, the end-user can post or retrieve observations. Fig. \ref{fig:Entity} shows an example of the Thing Entity used in the OGC SensorThings API. As seen, the code is simplistic, visible, and readable to any end-user.

SensorThings data, JSON, and metadata can be created, read, updated, and deleted via HTTP communications protocols. These protocols are commonly used by developers (POST, GET, PATCH, DELETE). They can easily and efficiently manipulate and create data exchanges between systems and networks. Each entity (as described above) has a unique ID for HTTP manipulation through RESTFUL API using a unique URL and query parameters to its respective server. Later, we will discuss several implementations that describe the deployment and testing of the OGC SensorThings (Fraunhofer Open Sensor Source, FROST) server and OGC SOS server (ISTSOS Server). Both these servers are open-source software that deploys the OGC standards.

The OGC SensorThings API can be used and managed by organizations that need to drive web-based platforms, share data, and provide analysis of IoT sensor data. Organizations that track the spread of the pandemic can quickly deploy and utilize these standards to give analysis and future trajectories of the pandemic. This analysis can aid governments in efficiently allocating and balancing resources where needed. The majority of IoT devices have proprietary software and APIs that are vendor-specific. The OGC STA can accelerate the deployment of any web-managed application or sensor without any disconcerted effort into translating other heterogeneous protocols or vendor-specific hardware or devices. The OGC STA can be embedded into any IoT hardware or software platform and interconnect with any OGC standard-compliant server strategically situated with a network. 

In summary, OGC SensorThings API offers the following benefits. 1) it permits the proliferation of new high-value services with the lower overhead of development and wider reach, 2) it lowers the risks, time, and cost across the whole IoT product cycle, and 3) it simplifies the connections between devices-to-devices and devices-to-applications \cite{21}.

\section{OGC SAT API Evaluation and Performance}\label{tab:sec_ogc}
Jazayeri et al wrong reference \cite{14}, it should be \cite{Jaz} has conducted extensive testing on the OGC SAT API to evaluate performance requirements when compared to other leading communications protocols and IoT network implementations. They have bench-marked OGC SAT efficiency and latency standards on class-1 IoT devices and compared results against other common devices and protocols. Additionally, the authors tested the memory utilization, occupation, the size of the HTTP request, the response length, and response latency. They have compared the OGC SAT API performance against class-1 IOT devices including PUCK over Bluetooth, TinySOS, and SOS over CoAP (Constrained Application Protocol). Before the evaluation of \cite{14}, a brief description of each of the IOT devices would be warranted.

PUCK over Bluetooth is a command protocol that utilizes a set of standards to access, read, and write data onto the memory. The PUCK over Bluetooth provides seamless interconnection of devices through serial cables or ethernet. The purpose of the PUCK, which does not handle external communication, is to provide a means of storing the sensor data in the memory. If the network utilizes the data, another protocol, e.g., Bluetooth, must be used to extract and place it on the network. Bluetooth will allow the sensor readings to be extracted and uploaded into a virtual repository. A sensor protocol is also needed to query the sensor capabilities and observation. PUCK over Bluetooth requires complex integration between the sensor, PUCK, and the network. Storing, extracting, and placing data on the network requires implementing a complex architecture. Accordingly, it does not provide a scalable solution. Finally, the authors compared performance results in three main categories, Memory management, Request Size and Response Length. For memory management, the experiment emphasizes the consequences of poor memory utilization and resource consumption. A Netduino Plus with an HTTP webserver was used to measure memory occupation and utilization for each protocol discussed above. They first measured the space the code occupied after code employment to the EEPROM of the Netduino Plus. According to the authors, the occupation of ROM served as a good indicator of the code's implementation complexity within ROM and RAM. Table \ref{tab:parameters} illustrates the results of the RAM and ROM memory occupation and complexity of the code.

\begin{table*}
\renewcommand*{\arraystretch}{1.4}
\caption{RAM and ROM Memory Occupation} 
\centering 
 {\begin{tabular}{|c c c c c c|} 
\hline
    & \textbf{Simple Web Service} & \textbf{PUCK over Bluetooth} & \textbf{TinySOS} & \textbf{SOS over CoAP } & \textbf{OGC SensorThings} \\  \hline 
    ROM (KB) & 16.08 & 8.48 & 11.72 & 29.13 & 26.11 \\
    RAM (KB) & 9.54 & 13.15 & 11.33 & 10.36 & 10.21 \\ \hline 
\end{tabular}}
\label{tab:parameters} 
\end{table*}

Research conducted by Bormann et al\cite{16} proposed a "tiny web service" and interface that is hosted in the application layers for providing "self-describable" and "self-contained" sensors. Each sensor provides a web interface to access the sensor's capabilities. The sensor device is independent of other nodes on the network to post or retrieve its data. This concept deemed the TinySOS (Sensor Observation Service), implements the OGC's Sensor Observation Service web interface platform used for retrieving sensor observations and raw metadata \cite{17}. The TinySOS is specifically developed for resource-constrained environments and applications. The implementation of this architecture as accomplished by \cite{15} supports a lightweight communications protocol, limited memory capacity, and limited memory utilization. Although the TinySOS can operate in resource-constrained environments, it utilizes a proprietary API to limit the device's ability to inter-operate within networks that host sensors from varying vendors or have other communications protocols. 

The final IOT OGC SOS implementation, studied in  \cite{14}, is the SOS over COAP. The Constrained Application Protocol is specifically used for constrained devices (nodes) to facilitate communications with the mainstream internet using similar protocols. CoAP is designed to communicate between devices on networks that are resource-constrained \cite{18}. The protocols utilize the fundamental features of HTTP communicating over constrained networks and simultaneously maintain a low overhead signature. The CoAP allows for machine-to-machine communications centered around the applications layer and implements Representational State Transfer (RESTful) services. The main difference between the CoAP and other protocols is using UDP (User Diagram Protocol) explicitly for speed and congestion control. In contrast, TCP (Transmission Control Protocol) uses congestion control algorithms that can cause unwanted retransmissions resulting in increased overhead and latency.

The network architecture of implementing SOS over CoAP requires the use of a proxy server strategically situated. These proxy servers allow the intercommunication and relay of sensor observations from node to node.CoAP can communicate between homogeneous devices and communications protocol but lacks compliance with the World Wide Web Standard \cite{19}. There have been efforts to integrate CoAP into the World Wide Web by wrapping the protocol into the Simple Object Access Protocol (SOAP). This fusion of protocols leads to undue complex messaging processing, drastically increasing latency and network overhead \cite{20}.

Table \ref{tab:parameters} also shows the RAM occupation comparison of all four protocols. Although some of the protocols performed well in ROM utilization and occupation, they proved inefficient in RAM utilization. The PUCK protocol was the worst in RAM utilization because of using a data transceiver and Bluetooth module. The OGC PUCK and the TinySOS also consumed a moderate amount of RAM due to their heavy reliance on parsing libraries for XML and JSON. SOS CoAP and OGC SAT had similar results on the lower end of the spectrum, with the OGC SAT performing slightly better than SOS CoAP. The best performer was the Simple Web Server which employs simplistic request validation and response generation protocols.

\subsection{Edge Level Implementation}
The design and implementation of a global COVID tracking system require the fusion of diverse technological advances made in AI, mobile devices, cloud computing, API, and web services such as the Geospatial Consortium. The system must take advantage of current AI models that can utilize ML to interpolate the infectious state of a person. The architecture requires the use of Mobile Edge Computing, which will provide or enhance the cloud computing capabilities at the edge of the system close to the tracking sensors \cite{6}. MEC is regarded as a key technology to provide users with low-latency, high services to these tracking sensors situated at the edge of the network. At the application layer, the Geospatial Consortium SensorThings API provides a framework that can be implemented to address interoperability issues.

To conceptualize the COVID tracking network, the underlying network layers, structural components, and corresponding technologies will be discussed. It is imperative to begin at the tactical edge where the IoT sensors and edge computing devices reside. These devices will be responsible for collecting, analyzing, and transferring relevant data to the next level of the architecture. Machine Learning (ML) has proven itself to be a prominent mechanism for modeling stats and predicting future occurrences. ML is used to solve real-world complex problems such as tracking COVID and providing live forecasting of the spread of the virus. \cite{7}.

Typically, IoT devices employed in the network will transfer all the collected data to a centralized computing system for storage and analysis. Such implementation, within the COVID tracking architecture, would consume valuable resources, such as memory and bandwidth, enabling high latency and an exponential increase in congestion. By using ML on edge computing devices, computational stresses and stresses on the network can be eased by leveraging the computational power of these devices. The research in \cite{8} demonstrates an 80 percent reduction of data flow within the network by implementing feature learning and deep learning models at the edge. ML Incorporated into MEC devices can ease congestion and stress on any multi-layered network.

\subsection{Open Geospatial Consortium as Cloud Services}
The Open Geospatial Consortium (OGC) is an international organization. In the OGC,  commercial, governmental, non-profit, and research organizations worldwide can cooperate on the implementation and deployment of geospatial content, web sensors, and IoT data.  The OGC has created agreed-upon standards for geospatial data. OGC services allow sensor data from participating organizations to be cataloged, retrieved, searched, and updated. The OGC has solved the issue of non-conforming communications protocols that are dependent on vendor-specific sensors. Using the current model, data sharing is severely impacted. The large amounts of data produced by these sensors are stored in varying digital metadata formats\cite{10}. Standardizing the metadata formats into a light weigh transport language will enable ease of sharing and digital translation of the data, increase network and data streaming efficiency, maximize bandwidth and allow the interoperability of vendor-specific sensors and IoT devices.

\subsection{Standardizing Digital Data}

In terms of standardizing metadata formats, this paper proposes the replication of using the Department of Defense standards of tracking mobile targets. These standards are unclassified, highly efficient, and lightweight in transmitting digital data \cite{12}. The metadata standards used by the Cursor on Target (developed by the MITRE organization) schema minimize the size of the data to bytes by using a uniform Extensible Markup Language (XML) with unique headers. This same protocol can be used for device-to-device communications within a structured network. The protocol can be used to track mobile patients as well as static patients within a hospital setting. In the next section, the proposed XML format and specific COVID tracking headers will be examined.

Standardizing digital data communications is critical in creating a "common language". Accordingly, all devices can report the data in a timely and efficient manner. This strategic approach would enable the use of any sensor despite the manufacturer and alleviate any vendor-specific device monopolizing the architecture. When the data is collected, the edge device will encapsulate the data in a specific schema common to all.  XML is a widely used markup language that is lightweight and used in any database or website for converting bytes into something readable. The bytes from the sensor are converted into ASCII code and inserted into the formatted XML schema, which would allow for convenient real-time information exchange. This Covid on Premises (CoP) (proposed protocol's name) schema would define mandatory fields for exchanging edge-level sensor data.

\begin{figure*}
    \centering
    \includegraphics[height = 7cm, width=10cm]{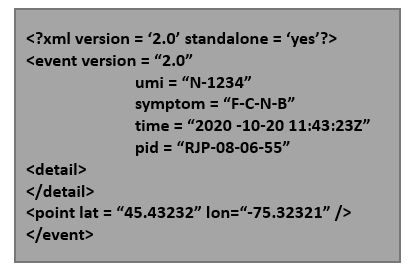}
    \caption{CoP Schema}
    \label{fig:Schema}
\end{figure*}

The XML schema would provide a framework and standard for data translation and exchange. The schema contains parent and child elements which can be easily transferred into databases containing SQL or JSON as their catalog language. This schema enables improved network performance, reduced power consumption, and creates an efficient interchange between endpoints. The schema would encapsulate minimum mandatory fields, shown in Figure \ref{fig:Schema}, as follows:

 \begin{itemize}
 \item [--] Unique Message Identifier (UMI) - this unique identifier would prevent the logging of duplicate data. It will act as the transmission's "serial number".
 \item [--] Symptom Identifier - for this identifier, the letters would be used corresponding directly to the symptom experienced. For example, if the sensor reported fever, cough, nausea, and perhaps loss of breath, the identifier would read as F-C-N-B. These symptom abbreviations would be cataloged into a database for easy recognition.
 
 \item [--] Time Stamp - very message would have a timestamp that will enable the data consumers to plot against time and allow for real-time analysis.
  
 \item [--] Patient Identifier (optional field) - The message would have the patient's name in the form of initials combined with the numeric date of birth. That would also prevent duplicate data within a specific time period
 
  \item [--] Geographic Location - every message will have the coordinates in the form of Lat / Lon of the geographic location from which the sensor has acquired the data.
 \end{itemize}

\subsection {Multi Access Edge Architecture}
The Covid Sensor Tracking System intends to automate the process of gathering intellectual data on the illness. This tracking system will enable medical professionals and local governments to obtain immediate analytical data to make informed decisions. The architecture should be large enough where most of the population is tracked through manual-reporting means such as medical staff or self-reporting within homes. Sensors can also be strategically placed within densely populated regions.

\begin{figure*}[ht!]
    \centering
    \includegraphics[width=1\textwidth,]{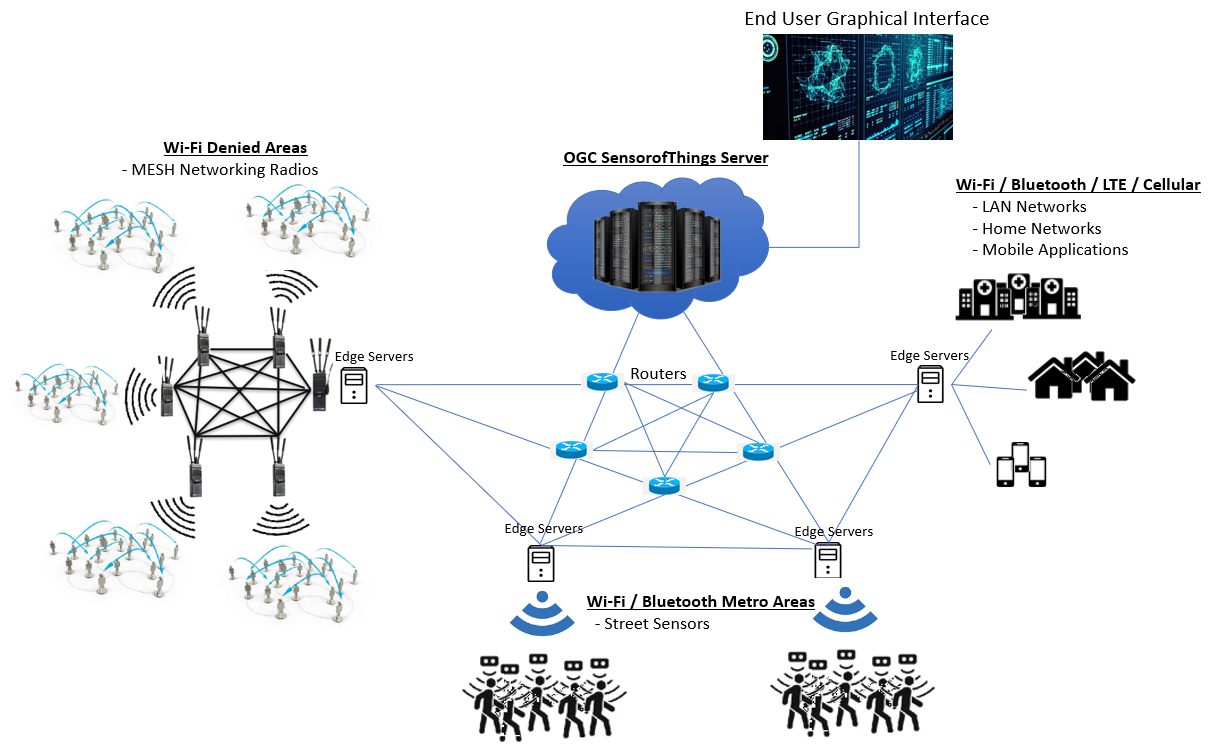}
    \caption{Network Diagram}
    \label{fig:Network}
\end{figure*}

This architecture implements Multi-Access Edge Computing with the addition of cloudlets to accomplish a new paradigm called Edge Mesh, proposed by  \cite{13}. The cloudlets are incorporated in such a manner that they act as the middle man between the terminal device and the originating devices. This layer which can be labeled as the abstraction layer, amongst other things, provides the computational resources for the edge devices. This layer may also have storage and routing capacity for network efficiency and forwarding. The architecture will allow for the prevention of traffic congestion, alleviates bandwidth, and prevents anyone device from depleting its computational and storage bearing resources. The network can share resources despite geographic location.

The computational process must become distributed across the networks. Distribution of computational power or sharing of computational power would allow low latency, optimal scalability, enhanced security, and prevent traffic congestion. The algorithmic and/or data filtering process would take place at the edge of the network versus transmitting the data to a centralized server or single processor. This organization would allow the efficient use of AI or Machine Learning algorithms.

As discussed in \cite {13}, the current Cloud computing paradigm suffers from four major issues, latency, security, privacy, and mobility, which have enabled architecture to move into the FoG computing paradigm. As defined by  \cite{14}, FoG computing is a "system-level horizontal architecture that distributes resources and services of computing, storage, control and networking anywhere along the continuum from cloud to Thing."

The Covid Tracking Network Architecture is distributed as follows:

 \begin{itemize}
 \item [--] OGC SensorThings API - At the Apex of the hierarchy, the OGC servers are provisioned as the consolidation point for all the sensor data. Data consumers would be able to download the analytical data to the graphical interface of preference. That would allow the user to represent the data in their required or desired format and/or representation.
 
 \item [--] End User Graphical Interface - The consumer can choose to view the data within the OGC website, view it as raw data, or download it onto their preferred graphical user interface for further processing or presentation.
 
 \item [--]  Edge Server - These servers are the core of the distributed shared processing. These servers, also called cloudlets, would share the computation processes amongst the network. They would be geographically dispersed and accessed according to how the data is routed amongst the network. If the resources of the local cloudlets are being consumed, the data will get transferred to the next cloudlets within the network.
 
 \item [--] MESH Networking Radios - These radios allow distribution of services within Wi-FI and Bluetooth denied areas. They can extend Wi-Fi or internet coverage in signal-denied areas. These radios are namely Mobile Adhoc Networks that can be deployed easily at relatively low cost when compared to establishing a communication infrastructure.
 
 \item [--]  Street Sensors - At the lowest level are the Street Sensors which operate within Wi-Fi-rich areas such as MAN networks in large cities. They would be deployed in and around chokepoints for automatic detection and collection of COVID metrics such as high body temperatures.
 
 \item [--]  LAN / Home Networks - These applications would be self and automated reporting applications such as high body temperature. The mobile tracking applications incorporation would contribute to the COVID tracking system, enabling home users to report any symptoms experienced.
 
 \item [--]  Core Routers - These routers are the backhaul of the network. The routers would enable the transfer of data in between cloudlets dependent on resource utilization.

 \end{itemize}

\section{OGC Server Implementation and Comparison}
This implementation of the OGC standards involves comparing the OGC SensorThings API server and OGC SOS server. In this experiment, the FROST Sever and the istSOS Server are used to represent the APEX of the COVID tracking network. Both standards on implemented on their respective servers and utilize autonomously incorporate and utilizes database management software, such as MySQL and PSQL. These servers relieve IoT devices from any computational and storage expenses.

The deployment of each server incorporated stand-alone docker containers, in which each instance had its image and dependencies. Furthermore, to test the efficiency of the protocol to include latency and CPU utilization, the respective servers were deployed to an Elastic Service Container provided by Amazon Web Services. Using AWS enabled expedited deployment to the World Wide Web and allowed for monitoring performance metrics such as CPU utilization, packet request, and latency. Each ECS instance is assigned an IP address for World Wide Web accessibility.

In this experiment, the goal was to produce a comparative analysis between the two API servers. Various metrics, such as packet size, latency, and CPU utilization were monitored, via AWS monitoring services. A Python script was used to reproduce the functions of sensors sending and requesting data to their respective servers and APIs. The Python script included a scalability feature in which sensor data was scaled up at specific intervals, starting from 1-1000 HTTP requests. These procedures were reproduced for the FROST-Server, incorporated the OGC SensorThigns API, and the istSOS Server, which incorporated the OGC SOS API.
 
 \begin{figure*}
\centering 
\includegraphics[height = 12cm, width=13 cm]{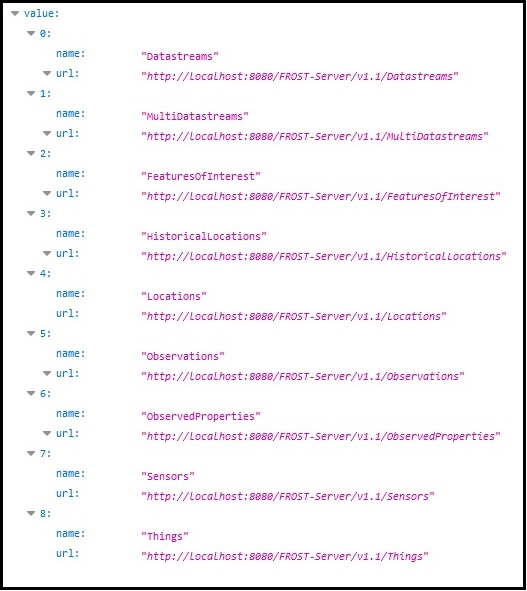}
\caption{FROST-Server API Navigation}
\label{fig:Navigation}
\end{figure*}

\begin{figure*}
\centerline{\includegraphics[height = 22cm, width=14cm]{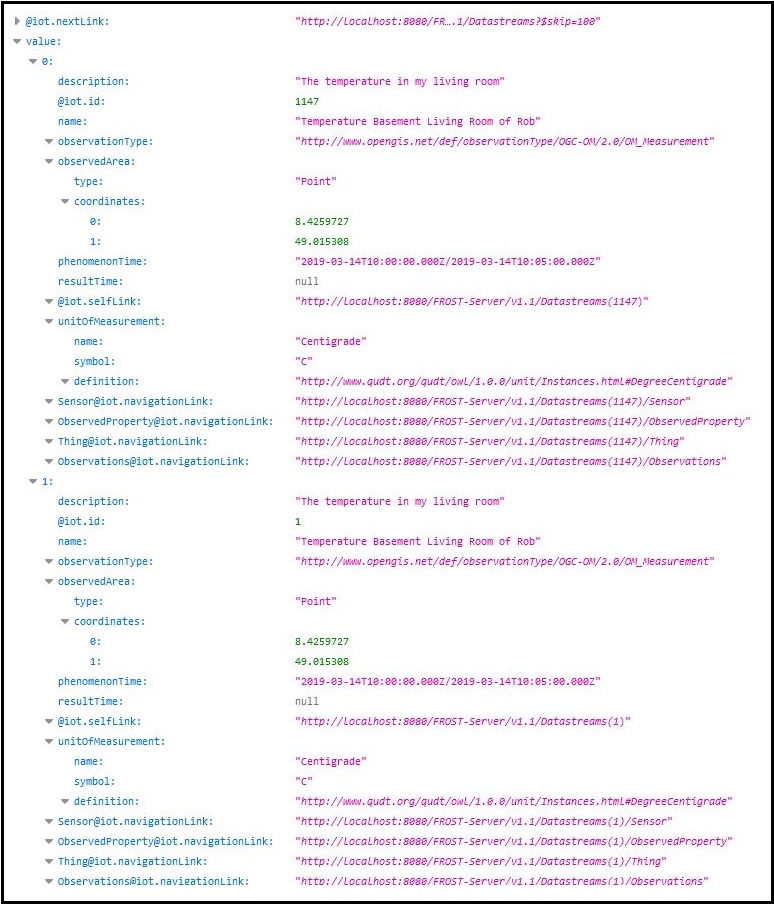}}
\caption{FROST-Server GET Response}
\label{fig:Response}
\end{figure*}

\subsection{OGC STA FROST-Server Selection}
The FROST-Server proved an efficient and expeditious means for deploying OGC SensorThings API standards. The FROST-Sever is an open-source platform that allows the end-user to quickly deploy the OGC STA API suite of protocols. The server can be implemented mainly using two deployment methods. The first is through a Docker container. The second is via Java servlets such as Tomcat, JETTY, or Wildfly. In this experiment, a Docker container was used to deploy the server. On the backend, the PostgreSQL database was used to store sensor information sent to the server.

The FROST-Server is packaged where the database proprietor can choose to use MQTT or HTTP as the transport protocol to and from the server for updating sensor data. It includes an all-in-one package where the transport protocol can be individually deployed. Since everything runs under the same JVM, the HTTP and MQTT entities can communicate directly. Therefore, substantial latency while executing requests will be eliminated. Furthermore, the FROST-Server is available in GitHub and packaged with extensive documentation on deployment and usage.

One of the main strengths of the SensorThings API is the ease of accessibility and navigation through the links that are produced as observations sent to the server. Each "thing" observation becomes an HTTP URL in which the user can click or cut and paste to traverse through the data and associations. The user can also use tools such as POSTMAN to retrieve these associations using simple POST and Get requests within the server. As shown in Figure \ref{fig:Navigation}, the FROST-Server handles the seven core entities of the SensorThings API. The server landing page provides the names of the SensorThings class available in conjunction with the URL of the location. The URL is in the form of a clickable link that directs the user to the storage location of that specific entity. Below is an illustration of the landing page. As illustrated in Figure \label{fig:Response}, Value objects can be accessed via the URL displayed on the landing page. Once the user clicks on the link or accesses the object via HTTP request, the SensorThings API will respond.

\section{OGC SOS Server Selection}

The OGC Sensor Observation Service standards are derived from the OGC Sensor Web Enablement (SWE) working group. The standards provide detailed definitions of the system that enables the utilization and development of Sensor-Web data. The standards define sensor technology, language syntax, and API employment architecture. Like many other standards, the OGC SOS provides a defined blueprint for establishing an environment where developers, end-users, and data providers can search, access, and process data from heterogeneous sensor networks across any geographic location.

\begin{figure*}
\centering
\includegraphics[height = 22cm, width=14cm]{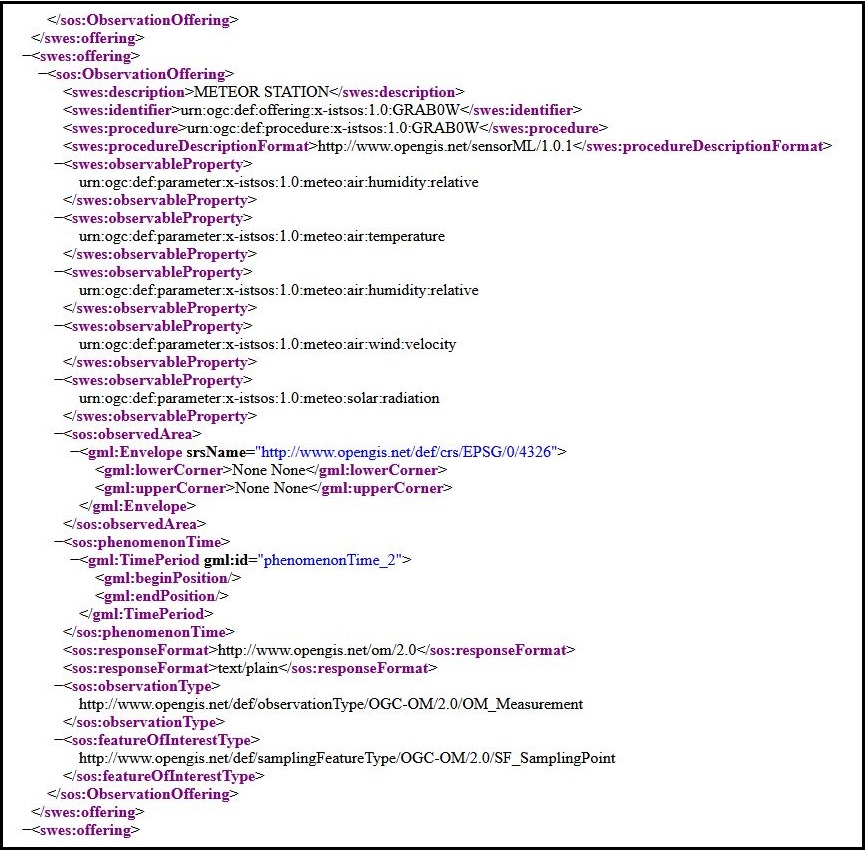}
\caption{istSOS GET Response}
\label{fig:GetResponse}
\end{figure*}

The SOS standards are established protocols that dictate the interactions between sensor observations, the server, and their transactional process. As shown in Figure \ref{fig:GetResponse}, like many other common APIs, the transaction occurs by using the HTTP protocol. The requests are sent, and responses are received using POST and GET methods, which are always XML compliant and JSON compliant. The OGC SOS implements at a minimum three of the main request entities. The other entities are optional and enhance the requested information. The SOS contains 5 core requests objects:
 \begin{itemize}
 \item [--] Observations - represents the values measured at any given timestamp and represented according to the OM standards data model.
 \item [--] Procedure - pertains to the data service provider or sensor that is sending the observation
 \item [--] Observed Properties - represents observed phenomena and is represented via a uniform resource identifier.
 \item [--] Features Of Interest - this category relates to the location of the sensor.
 \item [--] Offering - pertains to the collection of sensors used for grouping.
 \end{itemize} 

Istituto Scienze della Terra Sensor Observation Service (istSOS) is the implementation of the Sensor Observation Service standard from Open Geospatial Consortium by the Institute of Earth Sciences (IST, Istituto Scienze della Terra). istSOS provides a complete data management system that enables seamless interaction with data users and data providers. The istSOS is an OGC Sensor Observation Service server implementation written in the Python language. This project was initially designed for hydrologists managing observations at key oceanographic sites. 

The istSOS caters for the management of the OGC SOS standards that allow for managing observations from dynamic and static sensors. The istSOS OGC SOS server was chosen due to its extensive and comprehensive documentation regarding server deployment. In addition, the istSOS provides a detailed and vast tutorial with examples of registering sensors and get observations. The tutorial includes many examples with varying mock data from different types of sensors. That allows any user to expeditiously deploy the server with a flattened learning curve.


Furthermore, unlike other server implementation's, the istSOS OGC SOS server incorporates a well defined graphical user interface.The GUI incorporates every aspect of registering and monitoring a sensor including the usage of graphs and maps to plot and visualize sensor location data. The GUI allows an interface for implementing all facets of the SOS observation request standards. These facets include registering new sensors, updating metadata, verifying data, correcting and deleting data and the connection of a backend database. This simplistic yet powerful implementation of the OGC SOS standards via a GUI voids the need of the end user having any prior programmatic knowledge of any of the transport protocols such as MQTT or HTTP POST or GET requests. 

In order to utilize the POST and GET request, unlike the OGC STA, registering the observation is not a one instance approach. Within the OGC STA server, the user can register new sensor and observations using one instance of a HTTP POST. The OGS SOS protocol dictates that a user must first register a sensor and its intended procedures. After the sensor is properly registered including which data the user will update, only then can the user submit sensor readings and observations.

The istSOS data becomes modifiable for any given sensor observation when certain transactional operations are first executed. The first operation before insertion of observation data must be sensor registration and only then can the user insert an observation. The SOS operational commands are defined as follows: See Figure 8.
\begin{itemize}
    \item RegisterSensor - This is the initial executable command that allows the registration of a new sensor. SensorML encoding within an XML file is used to send the new registration service via HTTP POST request or can be done via istSOS GUI. See Figure 9.
    \item InsertObservation - This HTTP POST inserts observable data into the istSOS service using an XML file. The file specifies the identification of the registered sensor for proper classification. This action can also be accomplished via the istSOS GUI.
\end{itemize}
\smallbreak
\begin{figure*}[h!]
\centering
\includegraphics[width=12cm, height=10cm]{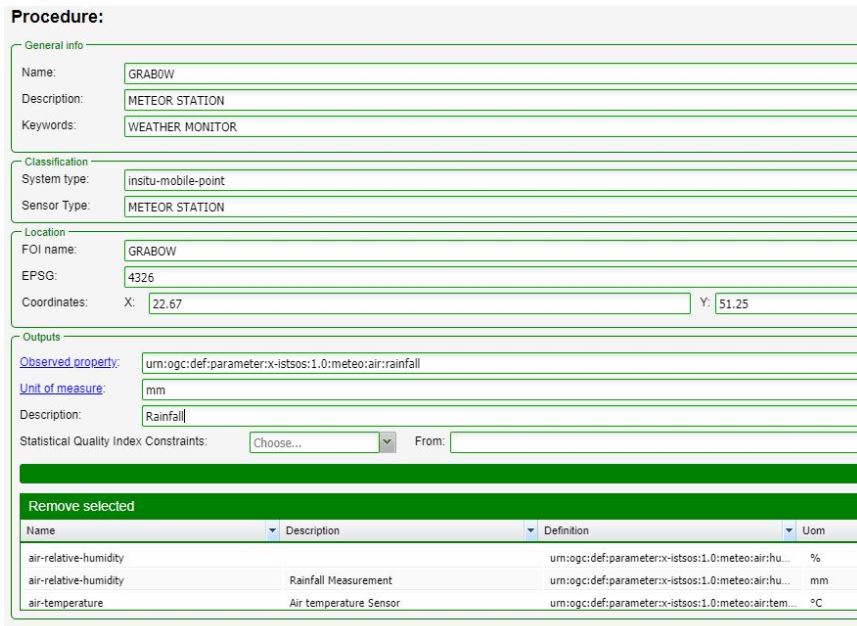}
\caption{istSOS Sensor Registration}
\label{fig:Registration}
\end{figure*}

One of the best features that encompasses the istSOS implementation is the ability to instantly graph, plot and visualize the data. Once the sensor data is uploaded, the user can instantly visualize the data in formats such as excel, csv or simply plotted on a map. This implementation of the istSOS server beautifully represents the ideas of a COVID tracking system using OGC standards. In an ideal situation, this type of implementation which includes the GUI and RESTFUL services could be also deployed using the OGC SensorThings API. This deployment encompasses the ideals of this paper in respect to networking aspects of the COVID tracking system. The FROST-Server demonstrates the minimum graphics to deploy the OGC SensorThings API, but lacks all the other features such as a graphical user interface, geographical plotting,and creating automated graphs. These features instinctively allows the user to obtain a higher level tactical knowledge of the trajectory of any pandemic. See Figure 11.

In addition to having the GUI available, the end user can decide to use RESTful services for HTTP POST, and GET request. The user can conveniently elect to either receive and send the data in JSON format or XML. This allows for flexibility amongst developers to incorporate the language of choice. Within the istSOS platform, the user has the convenience to POST and GET by accessing the Graphical User Interface or using HTTP. In this experiment, a python script was used to perform POST and GET request. This demonstrates the flexibility of accessing data within the istSOS. 
\begin{figure*}
\centering
\includegraphics[width=13cm, height=12cm]{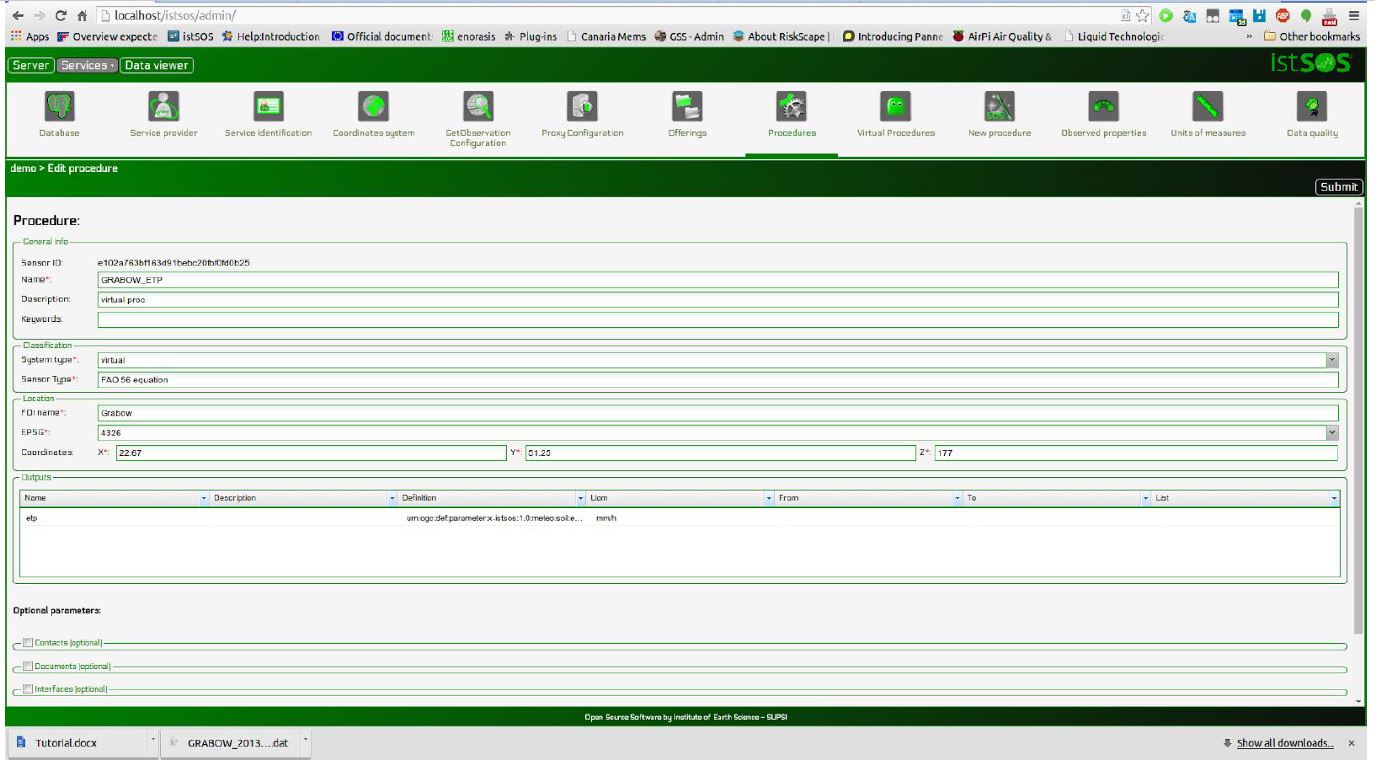}
\caption{istSOS GUI}
\label{fig:GUI}
\end{figure*}
istSOS allows for easily accessing and managing internal and external database repositories. Database Management is directly linked to the server and can also be accessed through HTTP or the GUI. The database is easily configurable, allowing the end-user to link an external database to the main server. The istSOS also allows the use of its default database system within a Docker container or file management system. Within this experiment, a POSTGRESQL database link was utilized. The POSTGRESQL database was configured with the servers URL and PORT (5432 by default) for seamless communication between server and database.This allowed database accessibility from the server or by logging into a terminal and directly accessing the sql database.

\begin{figure*}[h!]
\centering
\includegraphics[width=13cm, height=10cm]{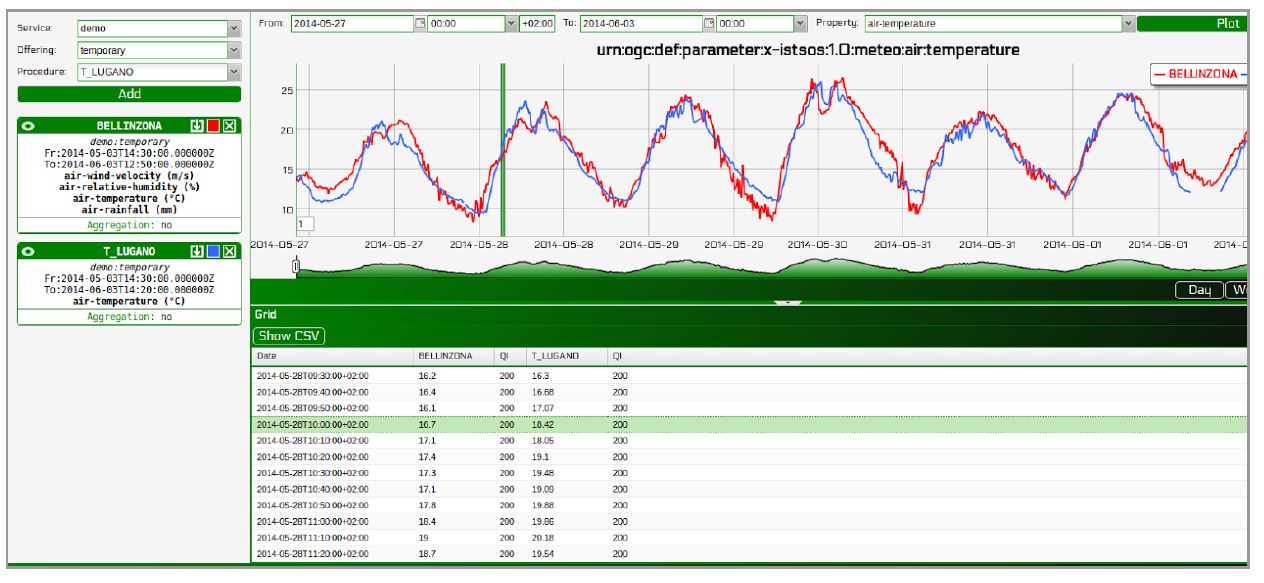}
\caption{istSOS Visualization}
\label{fig:Visualization}
\end{figure*}
\section{Quantitative Networking Results}
\subsection{Client Response Times}
This section contains networking results in respect to response times and CPU utilization when activating the HTTP requests. The maximum allowed requests at any given instance is limited to 1000. The graphs demonstrate a networking metric comparison between the istSOS server and the FROST-Server SensorThings API. Figure 13 shows the HTTP Response times of the HTTP queries. As expected the FROST-Server SensorThings API has a quicker response time than the istSOS Server. The FROST-Server is built on JAVA enabled server applications such as Java servlets. The FROST-Server employs the minimum architecture of a client-server model.

The istSOS response times are almost doubled in comparison to the FROST-Server. The istSOS's architecture is built on python. This is expected behavior since Java (which is utilized by the FROST-Server), is generally faster than python. Java is a compiled language, therefore making it much faster and more efficient than python. Python is an interpretive language and its human readable nature makes it faster to develop and deploy any application. As seen in the graph, the time response of istSOS is approximately doubled at any instance within the query size. The response times plateaued at about 700-800 Observations at about 700 milliseconds. The FROST-Server plateaued at about 350 milliseconds at approximately 800 observations. For single observations, the FROST-Server average 225 ms and the istSOS averaged 325 ms. As stated, this behavior was anticipated due to the python architecture of the istSOS.
\begin{figure*}[h!]
\centering
\includegraphics[width=13cm, height=10cm]{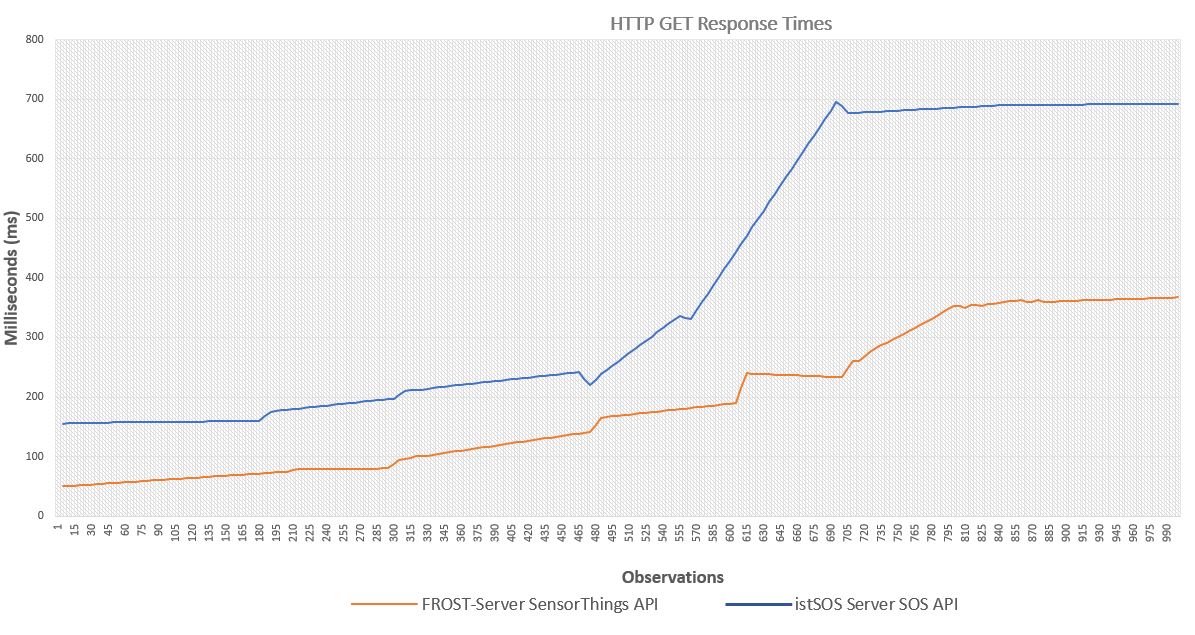}
\caption{Packet Size Comparison between istSOS and FROST-Server SensorThings API}
\label{fig:Packets}
\end{figure*}

\subsection{Client Response Sizes}
This section contains a comparative analysis of the request packet sizes when requesting an observation. Figure 13 illustrates the differences in packet sizes versus request queries. The observations request of the istSOS is the largest between the two implementations. The largest observation request resulted in 1400 kb at 1000 observations. Compared to the FROST-Server, the output size is almost double at 650 kb at 1000 observations. Both implementations have to the ability to reduce the packet sizes.This is accomplished by including resultFormat=dataArray when inserting a HTTP GET query.
\begin{figure*}[h!]
\centering
\includegraphics[width=13cm, height=10cm]{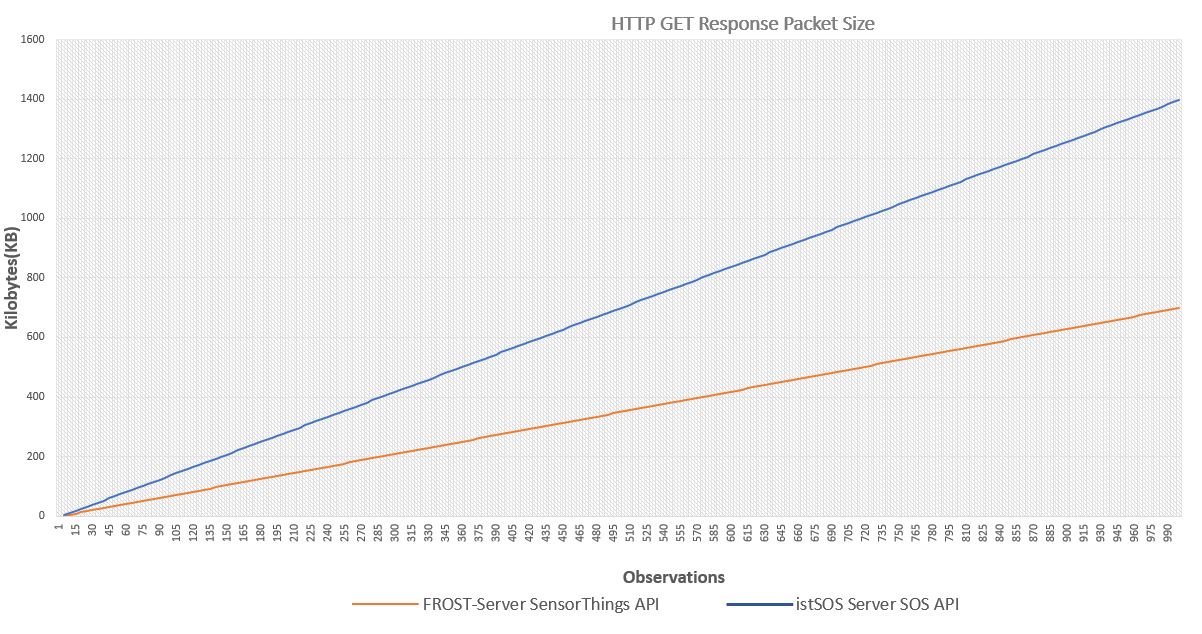}
\caption{Packet Size Comparison between istSOS and FROST-Server SensorThings API}
\label{fig:figure2}
\end{figure*}

The FROST-Server and istSOS server enables the user to select and utilize the Sensor Web Enablement DataArray format. This format eliminates many of the verbose features and allows the enduser to obtain only results and timestamps. This features substantially reduces the packet sizes by 50-60 percent in every instance and observation size. The SWE DataArray streamlines the requested data into two main categorical outputs:
\begin{itemize}
\item Metadata Values - timestamp, number of values, values encoding.
\item Sensor Value - Sensor metadata
\end{itemize}
 The FROST-Servers SensorThings API returned observations provide convenience in navigation. Each return observation attaches a link to the exact location, within the server, of any relational data. This link is in the form of an http hyperlink and by simply clicking on the link, the user can navigate to other related data. This also allows for smaller packet sizes since the end user obtains the data in form of hyper links versus obtaining all the data in one instance. The istSOS will return all the data within an XML file, including observations, timestamps and any other relational data within one file. See Figure 7.
\subsection{CPU Utilization Comparison}
The CPU utilization between the FROST-Server and the istSOS server showed noticeable difference during higher observation requests. The FROST-Server stayed constant relatively constant at approximately 15-20 percent utilization with the exceptions of a few spikes. The usage does not significant changes with increasing response sizes. The istSOS server showed higher CPU utilization percentages. The istSOS CPU usage remained at approximately 20-35 percent usage during the up to about 600 requests. Beyond 600 requests, the CPU usage jumped to about 45-50 percent, and remained within that area for the remainder of the observations up to 1000. Overall, the FROST-Server demonstrated lower CPU activity than the istSOS at all request intervals. The istSOS is built upon python and other components that make up the graphical user interface. Therefore, it would be expected behaviour of the istSOS to utilize more CPU. See Figure 14.
\begin{figure*}[h!]
\centering
\includegraphics[width=13cm, height=10cm]{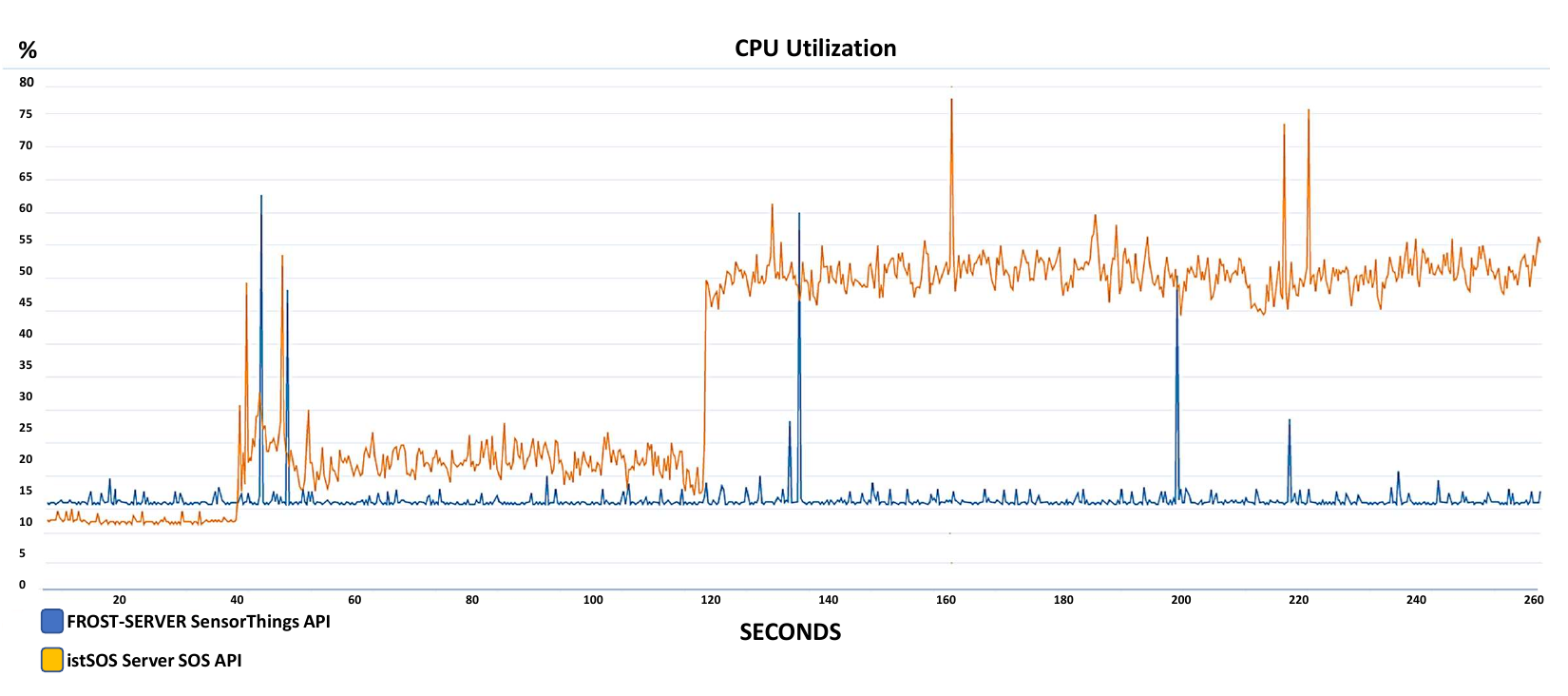}
\caption{CPU Utilization Comparison}
\label{fig:figure2}
\end{figure*}

\section {Conclusion}

 With the advent of the COVID-19 Pandemic, real-time tracking and data analytical has proved to be challenging using conventional methods.Containment of COVID-19 are directly affected by the following shortcomings, a) live monitoring of infected personnel b) transparency of data between infected nations c) lack of early analytical data including big data analytics and Artificial Intelligence d) late implementation of IOT devices for big data analytics \cite{2}. The lack of the aforementioned resources and/or implementations allowed COVID-19 to rapidly propagate the deadly virus globally. In turn, COVID-19 has effectively killed over a million people throughout the world.
 
 Deployment of a centralized tracking system would enable transparency and the potential to quickly establish the trajectory of any pandemic in the future. With a centralized tracking system local residents, governments and organizations such as the WHO (World Health Organization) could effortlessly track the infectious status of any region. This would allow residents to avoid areas of high statistical infectious rates. Governments could easily analyze and track the trajectory of the pandemic. This would allow organizations to prioritize containment efforts in geographic regions infected by the virus. Deployment of a centralized tracking system would also benefit health organizations by expeditiously pin pointing regions with immediate medical needs.
 
 One of the main technological issues of establishing a centralized tracking system is the diversity in IOT equipment and the sensor's reporting formality. It would be nearly impossible for every global region to employ the same vendor specific devices to obtain symptomatic COVID data pertaining to its residents. Further more, using a specific vendor would allow to monopolize certain markets and skew the real intent behind a centralized tracking system. 
 
 Establishing a centralized tracking system from scratch would prove to be a challenging effort. Nations will need to agree on a set of standards and protocols in which to communicate and share the data. Edge devices present the highest complexity of communication and sharing of data. Vendor monopolization of IOT devices will hinder the transparency and perceived trustworthiness of the data.
 
 Instead of "re-inventing the wheel" in terms of communications protocols and standard, the OGC standards has developed protocols that specifically cater to IOT devices and the formality of data sharing. With the development of the OGC SensorThings API protocols, a grand majority of the tracking system's networking gaps can be effectively filled. The OGC STA protocols allows for an effective communication and data sharing system between entities. The OGC STA standards eliminates the need to utilize vendor specific equipment. The standards will facilitate communications from any IOT edge device to a main stream server of the World Wide Web.
 
 In this paper, two of OGC API standards belonging to the Sensor Web Enablement protocols was compared. Each of the standards involved deploying its own open source API server implementation. The FROST-Server and the istSOS server was picked to demonstrate and analyze the performance metrics of the OGC SensorThings API and OGC SOS respectively.These servers allowed efficient and rapid deployment through AWS EC2 instances in the form of a Tomcat Server or Docker container. The AWS EC2 instance allows to deploy each of these servers on a Linux OS.
 
 Each of the implementations had exposed their strengths and weaknesses. In addition to evauluation of performance metrics when utilizing each of the API, subjective evaluations of each server's interface was also evaluated.
 
 The performance metrics obtained from OGC SensorThings API and OGC SOS API included Response Times, Packet Sizes and CPU utilization. In each of the performance categories, the FROST-Server OGC SensorThings API outperformed the OGC istSOS API server implementation. The performance outcome was expected due to different employment stacks used within the intricacies of the server development. The istSOS server front end is built upon JAVASCRIPT and the backend utilizes python, which has slower processing time and higher CPU utilization. The FROST-Server's backend is developed on java, which has faster processing time.
 
 Although, the FROST-Server outperformed the istSOS, in its current state, the FROST-Server lacks the robustness needed to integrate the plethera of platforms for the COVID tracking system. Users must be able to quickly and efficiently report data using various platforms such as mobile applications, sensor reporting and desktop applications. The FROST-Server only allows database access through the exclusive use of HTTP POST and GET requests. The FROST-Server does not employ a graphical user interface. In short, the user would need knowledge of HTTP protocols in order to report sensor data, data exchange or self-report data. The FROST-Server needs further development to accommodate access and reporting easability for the end-user.
 
 In contrast to the FROST-Server, the istSOS is based upon the OGC SOS API. The SOS API is not suitable for dynamic reporting of sensor data. The SOS API requires certain procedures (such as pre-registering a sensor and assigning specific measurement metrics) to be executed even before the user or sensor can freely share it's result. In addition, the SOS SensorML XML format proves too complex, and rigid. It lacks flexibility in reporting sensor simplistic sensor data. The SOS is better suited in reporting agricultural data as intended.
 
 Subjectively, the istSOS platform provided robustness and flexibility in regards to user interaction. The istSOS platform allowed the use of multiple avenues of data exchange. The most convenient avenue was utilization of the graphical user interface. The GUI makes available manual configuration of the platform. Manual configuration includes setting up the backend database, user profiles, request sizes and limits, reporting format, and the use of CSV files for extracting data and posting data. It also allows for manual registration and reporting of sensor observations.
 
 Perhaps the most intriguing feature of the istSOS platform is the instant visualization of the data. Once sufficient data is inserted into the database, the user can quickly visualize location information and statistical data on the platforms visualization tools. The platform will instantly graph data and plot the points on a map within the platform. This method of data integration and exchange would be better suited for the COVID tracking system due to its robust interface. The ideal platform would utilize the OGC SensorThings API in combination with istSOS's graphical user interface. The fusion of the two platforms (OGC SensorThings API and istSOS) will prove to be an efficient and exceptional for global inter-operability.

\bibliographystyle{IEEEtran}

\bibliography{bib}

\end{document}